\documentclass[journal=acsnano,manuscript=letter,layout=singlecolumn]{achemso}
\usepackage[utf8]{inputenc}
\usepackage[version=3]{mhchem} 
\usepackage{graphicx}
\usepackage{color}
\usepackage{epstopdf}
\usepackage{here}
\usepackage[utf8]{inputenc}
\usepackage{amsthm}
\usepackage{amsmath}
\usepackage{amssymb}
\usepackage{esint}
\usepackage{graphicx}
\usepackage{upgreek}
\usepackage{xspace}
\usepackage[normalem]{ulem}
\usepackage{nicefrac}

\newcommand{\didv}{$dI/dV$}

\usepackage{soul}
\usepackage{braket}
\usepackage[backref=false,
bookmarksnumbered=true,
bookmarks=true,
bookmarksopen=true,
colorlinks=true,
citecolor=blue,
linkcolor=blue,
anchorcolor=green,
urlcolor=blue,unicode=false]{hyperref}

\let\baraccent=\= 
\renewcommand{\=}[1]{\stackrel{#1}{=}} 

\newcommand{\Fig}[1]{Fig.~\ref{fig:#1}}
\newcommand{\Figure}[1]{Figure~\ref{fig:#1}}

\DeclareMathOperator{\unitspace}{\,}

\author{Benjamin\ W.\ Heinrich}
\affiliation{Fachbereich Physik, Freie Universit\"at Berlin, Arnimallee 14, 14195 Berlin, Germany} 
\author{Christopher Ehlert}
\affiliation{Institute of Chemistry, Universit\"at Potsdam, Karl-Liebknecht-Strasse 24--25, 14476 Potsdam, Germany} 
\alsoaffiliation{Department of Chemistry, Wilfrid Laurier University, 75 University Ave W, Waterloo, ON N2L3C5, Canada}
\author{Nino Hatter}
\affiliation{Fachbereich Physik, Freie Universit\"at Berlin, Arnimallee 14, 14195 Berlin, Germany} 
\author{Lukas Braun}
\affiliation{Fachbereich Physik, Freie Universit\"at Berlin, Arnimallee 14, 14195 Berlin, Germany} 
\author{Christian\ Lotze}
\affiliation{Fachbereich Physik, Freie Universit\"at Berlin, Arnimallee 14, 14195 Berlin, Germany} 
\author{Peter Saalfrank}
\affiliation{Institute of Chemistry, Universit\"at Potsdam, Karl-Liebknecht-Strasse 24--25, 14476 Potsdam, Germany} 
\email{peter.saalfrank@uni-potsdam.de}
\author{Katharina\ J.\ Franke}
\email{franke@physik.fu-berlin.de}
\affiliation{Fachbereich Physik, Freie Universit\"at Berlin, Arnimallee 14, 14195 Berlin, Germany} 

\date{\today}

\title{Control of oxidation and spin state in a single-molecule junction}
\begin{document}

\begin{abstract} 
The oxidation and spin state of a metal-organic molecule determine its chemical reactivity and magnetic properties. Here, we demonstrate the reversible control of the oxidation and spin state in a single Fe-porphyrin molecule in the force field of the tip of a scanning tunneling microscope. Within the regimes of half-integer and integer spin state, we can further track the evolution of the magnetocrystalline anisotropy. Our experimental results are  corroborated by density functional theory and wave function theory. This combined analysis allows us to draw a complete picture of the molecular states over a large range of intramolecular deformations.

\end{abstract}

\maketitle 

The oxidation state of an atom describes the number of its valence electrons in a very simple picture. Within this framework, redox reactions are understood as changes of the oxidation state by integer numbers (of charges). In metal-organic complexes,  such reactions typically involve a change in the number of bonds to neighbouring atoms. They are particularly important in nature, where they occur, e.g., in the oxygen metabolism by O$_2$ coordination to haemoglobin. Furthermore, they are important in the perspective of spintronics, because an oxidation reaction also involves a change of the spin state.

Surface-anchored metalloporphyrins are model systems, where the interplay between ligand geometry, oxidation state, magnetism and (spin) transport has been studied in great detail down to the single molecule level \cite{Auwarter2015,Gottfried2015,Wende2007,Iacovita2008,Tsukahara2009,Wackerlin2010,Schmaus2011,Wackerlin2012,Hermanns2012,Jacobson2017}. Scanning tunneling microscopy and spectroscopy (STM/STS) has been used to show that additional ligands like H, CO or Cl can induce changes in the oxidation and spin state of the core metal ion~\cite{Gopakumar2012JACS,Strozecka2012,Liu2013,Heinrich2015Ani}. Most STM experiments assume that the tip is a non-invasive tool in the investigation. However, the STM tip can perturb the molecular properties when it is brought in sufficiently close proximity to the molecule. As such, the coupling of a molecule to the surface can be modified, which can lead to the appearance/disappearance of a Kondo resonance in case of magnetic molecules~\cite{Hiraoka2017,Abufager2017}. Similar changes had previously been only detected in two-terminal break junction experiments, where the stress upon opening/closing the junction is assumed to be very large~\cite{Parks2010}. Recently, it has been shown that far before contact formation, an STM tip may already affect intramolecular properties. In the far tunneling regime, the forces exerted by the STM tip are already sufficiently large to induce small conformational changes within the molecule, which affect the crystal field splitting of the central ion, and, hence, its magnetic anisotropy~\cite{Heinrich2015Ani}.

Here, we show that the tunneling regime bears even more degrees of freedom, which can be reversibly controlled by the STM tip. 
For this purpose, we adsorb iron(III) octaethylporphyrin chloride (Fe-OEP-Cl) on a Pb(111) surface.
In the absence of the tip, the d$^5$ configuration of Fe$^\mathrm{III}$ gives rise to an $S=5/2$ spin state in the ligand field.
We make use of the possibility to controllably and reversibly manipulate the Fe--Cl bond by the force field of an STM tip. We show that the forces do not only lead to a change in the magnetic anisotropy as was reported previously\cite{Heinrich2013}, but that they can also be used to induce a reversible redox reaction, implying the reversible change from a half-integer to an integer spin state. Both states are not affected by sizable scattering from substrate electrons, i.e., their spin state is not quenched by the Kondo effect. In combination with the use of superconducting electrodes, which increase the excited state spin lifetime compared to metal electrodes\cite{Heinrich2013}, this renders this system suitable for potential spintronic  applications. 

The experiments are supported by theoretical simulations. By means of Density Functional Theory (DFT) and Wavefunction Theory (WFT) calculations, we determine the relative energies of different spin manifolds of Fe-OEP-Cl as a function of the Fe--Cl distance. Our simulations reproduce the experimentally observed changes in the magnetocrystalline  anisotropy and indicate that at large Fe--Cl distances, the integer $S=2$ state is favored.

\section{Results and Discussion}
\begin{figure}[t]
  \includegraphics [width=0.7\textwidth,clip=]{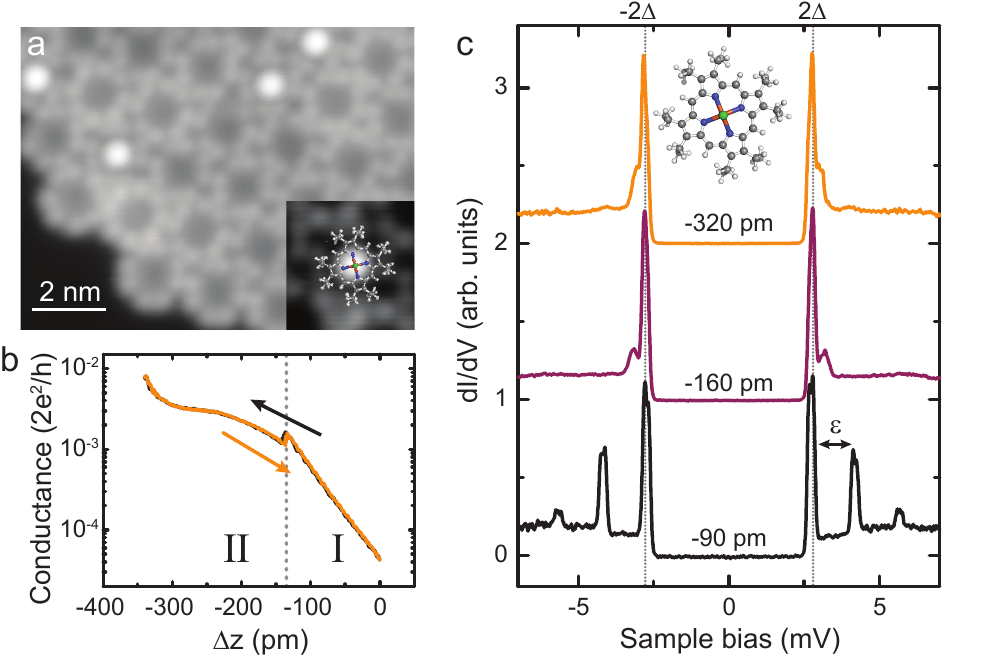}
  \caption{Fe-OEP-Cl on Pb(111). (a) Topography of a monolayer island. Two types of molecules are observed: Fe-OEP (dim center) and Fe-OEP-Cl (bright center). (b) Conductance at 10\,mV vs. tip-sample distance $\Delta z$ measured with the tip place above the center of a Fe-OEP-Cl molecule. A sudden drop in the conductance is observed at $\Delta z \approx -130$\,pm.  (c) $dI/dV$ spectra at different tip-sample distances  $\Delta z$ show signatures of inelastic excitations at different energies.}
\label{FigIz}
\end{figure}

We deposited Fe-OEP-Cl molecules on an atomically clean Pb(111) surface held at $\approx 240\,$K. This leads to self-assembled mixed islands of iron-octaethylporphyrin-chloride (Fe-OEP-Cl) and iron-octaethylporphyrin (Fe-OEP) as can be seen in the STM images obtained at low temperature (Fig.~\ref{FigIz}a). The individual molecules can be identified by the eight ethyl legs. The chlorinated species appears with a bright protrusion in the center, whereas the dechlorinated species appears with an apparent depression in the center. The dechlorination process  is probably catalytically activated by the surface and occurs  upon adsorption. 
The magnetic excitation spectra of both species have been characterized previously\cite{Heinrich2013,Heinrich2015Ani}. 
As mentioned above, in Fe-OEP-Cl the Fe center lies in a $+3$ oxidation state with a spin of $S=5/2$. Dechlorination changes the oxidation state to $+2$ with the resulting spin state being $S=1$.
 This change of oxidation and spin state is naturally expected when the Fe center is subject to such a drastic modification as removing one ligand. It was also observed that the STM tip can have an influence on both individual species. The Fe ions are subject to crystal fields, which split the $d$ levels and induce a sizable magnetocrystalline anisotropy and, hence, a zero field splitting (ZFS). By approaching the tip to the molecule, for both species the ZFS has been shown to be modified by the local potential of the tip\cite{Heinrich2015Ani}.

Here, we explore if the tip potential can be used to change the local energy landscape even further such that a reversible change not only of the ZFS, but also of the spin state takes place. For this, we target at the Fe-OEP-Cl molecule because it bears more flexibility with its additional ligand than the dechlorinated species. We first explore the change of junction conductance upon approaching the tip on top of the Fe-OEP-Cl molecule in Figure~\ref{FigIz}b.  After a first exponential increase of the conductance with reduced distance ($\Delta z<0$), which is characteristic for the tunneling regime (regime I), there is a reversible drop in the conductance in the order of 30\% of the total signal.  This can be linked to a sudden change of the junction geometry and/or the electronic structure of the complex. Further decreasing the distance increases the conductance again, yet with a smaller and decreasing slope (regime II). At $\Delta z\approx-330$\,pm, the conductance increases again with increasing slope. 
The different slopes indicate different structural/ electronic configurations of the molecule in the junction. To characterize the magnetic properties of the molecule in these regimes, we record \didv\ spectra at different $\Delta z$ (Fig.~\ref{FigIz}c).  In all spectra, we observe an excitation gap with a pair of quasiparticle excitation peaks at $\pm (\Delta_{tip}+\Delta_{sample})$, with $\Delta_{tip}$ and $\Delta_{sample}$ being the real part of the order parameter of the superconducting tip and sample, respectively. 
In regime I ($\Delta z=-90$\,pm), there are two additional pairs of resonances outside the gap. These resonances are due to inelastic spin excitations on a superconductor \cite{Heinrich2013,Fransson2015}. Replicas of the quasiparticle resonances appear at the threshold energies for the opening of an inelastic tunneling channel. The first inelastic peak corresponds to a transition between the
$\mathrm{M}_{\mathrm{s}}={\pm{1}/{2}}$ to  the $\mathrm{M}_{\mathrm{s}}={\pm{3}/{2}}$ state within the $S=5/2$ manifold. The second excitation from the $\mathrm{M}_{\mathrm{s}}={\pm{3}/{2}}$ state to the $\mathrm{M}_{\mathrm{s}}={\pm{5}/{2}}$ state becomes possible at large current densities due to spin pumping~\cite{Heinrich2013}. 
In regime II, i.e., after the sudden change in the molecular conductance, both inelastic excitations vanish. Instead, an excitation of lower energy is observed close to the superconducting gap edge (see, e.g., the spectrum at $\Delta z=-160$\,pm). At even smaller tip--molecule distance (exemplary spectrum at $\Delta z=-320$\,pm), we observe an inelastic excitation of even lower energy, which merges with the coherence peak.

\begin{figure} [t]
  \includegraphics [width=0.7\textwidth,clip=]{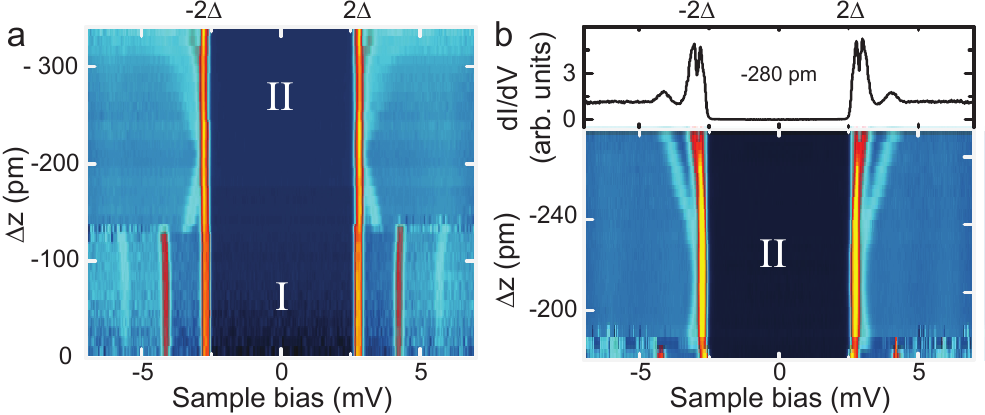}
  \caption{(a) Pseudo-2D color plot of $dI/dV$ spectra of the molecule in Fig.\,\ref{FigIz}b and \ref{FigIz}c as a function of $\Delta z$. (b) Pseudo-2D color plot of $dI/dV$ spectra of a different molecule with a different tip. $\Delta z$ increment: 5\,pm. The spectrum at $\Delta z=-280$\,pm is shown on top and evidences a low-energy excitation close to the superconducting gap edge. For additional data on different molecules and with different tips see SI. }
\label{Fig2Dplot}
\end{figure}

To track the evolution of the inelastic excitations we record a series of spectra while approaching the STM tip in small intervals. A 2D color plot of these spectra is shown in Fig.~\ref{Fig2Dplot}a. Within regime I, we observe a shift of both excitations to larger energies upon tip approach as a result of an increased ZFS. This has been  qualitatively explained as a response to a distortion of the ligand field~\cite{Heinrich2015Ani}. The sudden drop in conductance at the transition to regime II is also reflected in an abrupt change of the \didv spectra. The two excitations of the $S=5/2$ manifold disappear and a new excitation appears. It shifts to lower energies and merges with the quasiparticle excitation peaks when approaching the STM tip to $\Delta z\approx -200$\,pm. Eventually, a faint excitation emerges from the quasiparticle excitation resonances for $\Delta z< -220$\,pm and a second excitation departs from the gap edge at $\Delta z\approx -290$\,pm.  A high-resolution 2D plot of regime II (for a different molecule) is shown in Fig.~\ref{Fig2Dplot}b. It resolves the latter two resonances more clearly (note that their appearance is shifted in $\Delta z$ with respect to the molecule shown in  Fig.~\ref{Fig2Dplot}a).

\begin{figure}[t]
  \includegraphics [width=0.7\textwidth,clip=]{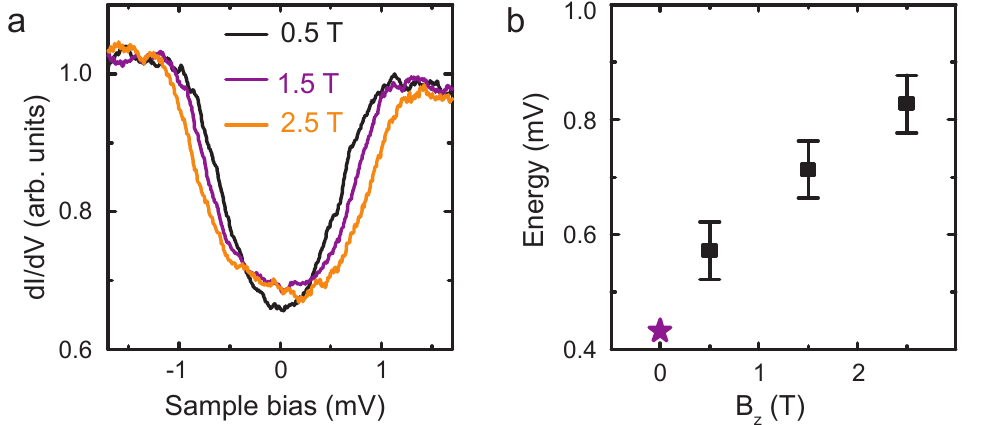}
  \caption{(a) $B$-field dependent $dI/dV$ spectra at $\Delta z=-150\,$pm. A shift of the excitation step to higher energies is observed with increasing $B$ field. (b) Step energies as a function of $B$-field (black squares). 
The purple asterisk indicates the energy as extracted from the spectrum recorded with tip and sample in the superconducting state. The excitation shifts to higher energies with increasing $B$ field.}
\label{FigZeeman}
\end{figure}

Whereas the origin of the two excitations in regime I is already understood,  regime II was not explored in earlier experiments. We note that regime II does not indicate the desorption of the Cl ligand, because Fe-OEP shows characteristic spin excitations at larger energies \cite{Heinrich2015Ani}, which are absent here (see Fig.\,S2 of the SI). Furthermore, the junctions are stable: when we retract the tip, we reversibly enter into regime I. Indeed, we can precisely and reversibly tune the junction conductance and, hence, the energy of the excitations.  We first determine whether the new excitations are of magnetic origin. By applying an external magnetic field perpendicular to the sample surface, superconductivity in substrate and tip are quenched. Here, inelastic excitations give rise to steps in the \didv spectra at the excitation threshold as expected for a normal metal substrate (Fig.~\ref{FigZeeman}a). Importantly, we observe a shift of the steps to larger energies in response to an increasing external field. The extracted step energies agree with a Zeeman shift (Fig.~\ref{FigZeeman}b), which evidences their magnetic origin.

\begin{figure}[t]
  \includegraphics [width=0.7\textwidth,clip=]{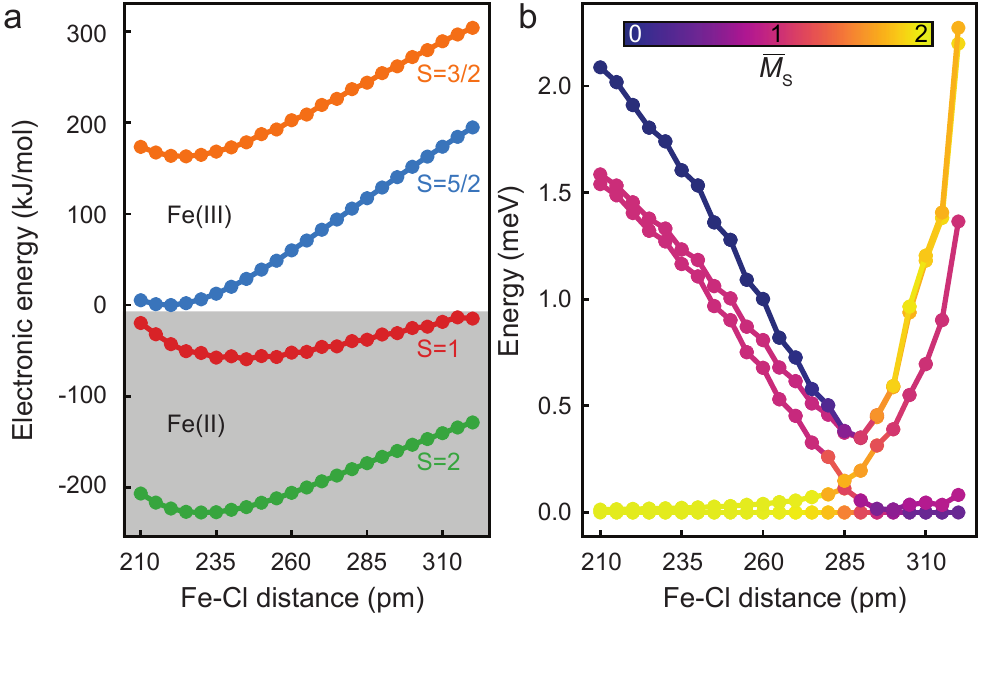}
  \caption{(a) Potential energies of different spin manifolds as a function of Fe--Cl distance,  determined 
 by a combination of DFT optimization and WFT single-point calculations (see text). Half-integer and integer $S$ values 
 refer to the neutral (Fe$^{\mathrm{III}}$) and 
 anionic (Fe$^{\mathrm{II}}$) complexes, respectively.
  (b) ZFS of the $S=2$ manifold. The colors indicate contributions of the eigenstates of the spin projection operator. We calculated the average ($\bar{M}_s$) as the sum of the absolute eigenvalues multiplied by the weights of contributing eigenstates. Note that at 210\,pm the lowest two eigenstates are almost degenerate, while for Fe--Cl distances larger than 285\,pm, the two highest states are almost degenerate. 
}
\label{Fig_theory}
\end{figure}


Summarizing the experimental results, there is evidence for a sharp transition from the $S=5/2$ state to a different spin state when approaching the STM tip on top of Fe-OEP-Cl. In principle, this could either be a spin crossover to another half-integer spin state or a change to an integer spin state, which necessarily needs to be accompanied by a change in oxidation state. In order to gain further insights, we performed DFT calculations using the B3LYP exchange-correlation functional\cite{Becke1993} as well as WFT based simulations in the form  of the so-called CASSCF/NEVPT2 method\cite{Malmqvist1989,Angeli2001,Angeli2001a,Angeli2002,Angeli2006} according to Refs.\cite{Atanasov2011,Stavretis2015}. Details are described in the Methods section and in the SI.
We assume that the tip-molecule interaction, 
in the sense of a Lennard-Jones potential, is attractive at not too short tip-molecule distances. 
 Upon approaching with the STM tip, the Cl ion will thus be attracted by the tip. We model the effect of such a potential in an isolated Fe-porphin molecule by increasing the Fe--Cl distance. In porphin, the eight ethyl groups are replaced by H atoms. We have previously checked that the ethyl legs do not affect the electronic and magnetic properties of the complex (see Fig. S6 in the SI).

We used DFT optimization to compute the relaxed structure while scanning and keeping the Fe--Cl distance fixed in three different spin states of the neutral Fe(III)-complex, i.e., with spin quantum numbers $S=5/2$; $S=3/2$; $S=1/2$. Based on these results, we calculated the potential energies by the CASSCF/NEVPT2 method at the DFT geometries (Fig.~\ref{Fig_theory}a; corresponding {\it pure} DFT potential curves are shown in Fig. S8 of the SI). In agreement with previous experiments and calculations, we find that the ground state exhibits a spin state of $S=5/2$. Here, the Fe--Cl distance amounts to \mbox{220\,pm}. This configuration represents the unperturbed molecule on the surface, i.e., when the tip is far away. 
While the energy difference between the $S=5/2$ and $S=3/2$ states decreases, the latter remains at larger energies throughout the whole range of reasonable Fe--Cl distances. Hence, a spin crossover scenario as an explanation for the sudden change in experimental spin excitations from regime I to regime II can be excluded\footnote{We can also experimentally exclude such a spin crossover, because the appearance of two excitations in regime II cannot be captured in a $S=3/2$ system.}. The $S=1/2$ spin state lies even higher in energy and can therefore be excluded to play a role as well.

Hence, we also calculated the potential energies of the anionic Fe(II)-complex in its $S=2$ and $S=1$ spin states\footnote{The $S=0$ state can be disregarded as it would not show any spin excitations.}. The $S=2$ state is found at significantly lower energies than the $S=1$ state. Within increasing Fe--Cl bond length, the potential energy increases but not with the same slope as in the $S=5/2$ state. As a result, the 
 energy difference between $S=5/2$ and $S=2$, i.e., the electron affinity 
 of the neutral complex {\it increases} with increasing Fe--Cl distance. 
 Together with a resulting image charge stabilization of the anionic complex and a 
 relatively low work function of Pb of about 4\,eV, this is the main driving 
 force for an electron transfer  from the tip or substrate to the molecule 
 at a certain critical, elongated Fe--Cl distance.

To check whether the experimentally observed spin excitations in regime II match with this scenario, we calculated the ZFS of the anionic system. The results are shown in Fig. \ref{Fig_theory}b. Overall it is observed that the state energies are very sensitive with respect to geometrical variation. The discussion of this behavior can be split in two parts: a first part (i) with Fe--Cl bond length shorter than \mbox{285\,pm} and a second part (ii) with Fe--Cl bond lengths longer that \mbox{285\,pm}. 

In part (i) the ground state is composed of two states, which are almost degenerated. The eigenvalues of the spin projection operator on the $z$ axis have mainly $M_s=\pm 2$ character. The third and fourth state have eigenvalues of $M_s = \pm 1$, while the fifth state has the $M_s = 0$ eigenvalue. The splitting can be easily rationalized with the  
 help of an effective Hamiltonian method \cite{Ganyushin2006} and related to 
 so-called axial and rhombic ZFS parameters $D$ and $E$, respectively, as explained in the SI.

With increasing bond length the third, fourth and fifth state decrease in energy and mixing occurs between the states.
Inelastic electron tunneling requires $\Delta M_s=0,\pm 1$. Hence, one can observe excitations from the $M_s=\pm 2$ ground states to the $M_s = \pm 1$ states. The energy of this excitation is in the order of 1\,meV with decreasing energy until the bond length amounts to \mbox{285\,pm}. This behavior of one excitation with decreasing energy is found in experiment at the initial stages of regime II, i.e., right after the spin transition.

When the Fe--Cl bond length becomes longer than \mbox{285\,pm} [part (ii)], the lowest two states are again almost degenerate, but have changed their character. Both states are now mixtures of several eigenvectors of the spin projection operator. The first one is dominated by $M_s = 0$, the second by $M_s =  \pm 1$. The third state is mainly of $M_s = \pm 1$ character, while the fourth and fifth are dominated by $M_s = \pm 2 $. The energy of the three highest states increase now monotonously with elongated Fe--Cl bond lengths. Since the nearly degenerate ground state has $M_s = \pm 1$ and $M_s = $0 character, inelastic excitations are allowed to both the $M_s = \pm 1$ and $M_s = \pm 2 $. These two excitations are indeed found in experiment in the later part of regime II. In agreement with theory, their energies increase with increased Fe--Cl bond length. The lower intensity of the second excitation in experiment may be ascribed to an underestimated energy splitting of the $M_s = \pm 1$ and $M_s = $0 state in the calculations. A finite splitting would require the second state to be thermally occupied in order to be able to observe an inelastic transition and thus be reduced in intensity.

In conclusion, we have established a fully reversible switching of the spin and oxidation state in a single molecule including the possibility to fine-tune the magnetocrystalline anisotropy. By approaching the STM tip on a Fe-OEP-Cl molecule, we have observed drastic changes in the spin excitation spectra. The force exerted by the STM tip leads to a deformation of the molecule, in particular to an elongation of the Fe--Cl bond length. Initially, this affects the magnetocrystalline anisotropy of the $S=5/2$ state. At a critical bond length, a sudden change of oxidation and spin state occurs, which leads to a distinctly different fingerprint in the excitation spectra. Our combined DFT and CASSCF/NEVPT2 simulations allow for an unambiguous identification of the experimentally observed excitations. Their behavior under the force field agrees well with the experiment. Our experiments provide an intriguing method for a reversible and controlled way of changing not only the magnetic anisotropy, but also and more drastically, the spin state. They open new pathways for controlling magnetic properties by external forces.

\section{Methods} 
All experiments were performed in a SPECS JT-STM, a commercial  low-temperature scanning tunneling microscope (STM) working under UHV conditions at a base temperature of 1.2\,K. We obtained clean and superconducting surfaces of the Pb(111) single crystal by cycles of Ne$^+$ ion sputtering and subsequent annealing to $\approx 430\,$K. 
In order to circumvent the Fermi-Dirac limit in energy resolution, superconducting, Pb-covered W tips were used, which allow us to improve the energy resolution down to $\approx 60\,\mu$V when employing elaborated RF-filtering and grounding schemes~\cite{Ruby2015Pb}.
All spectra are recorded in constant-height mode with an offset $\Delta z$  (as indicated in the figures) in the tip--sample distance relative to the set point: $U=50\,$mV; $I=200$\,pA.  
The differential conductance $dI/dV$ is measured 
using standard lock-in technique with a modulation frequency $f=912$\,Hz and a voltage modulation $U_{mod}=35\,\mu$V$_{rms}$.

We used the ORCA (version 3.0.3) program package \cite{Neese2011} to optimize geometries and to calculate potential energy surface scans along the Fe--Cl distance for several 
 spin multiplicities. The B3LYP density functional \cite{Becke1993} in combination with the def2-TZVP basis set \cite{Weigend2005} was utilized.

For the geometries along the potential energy scans, the Zero-Field parameters have been computed by using a 
correlated WFT methodology\cite{Atanasov2015,Ganyushin2006} based on the complete active space self-consistent field method (CASSCF)\cite{Malmqvist1989}. In order to take into account dynamical correlation, the NEVPT2 \cite{Angeli2001,Angeli2001a,Angeli2002,Angeli2006} method has been used on top of CASSCF. 
 
Subsequently, the effects of spin-orbit and spin-spin interactions were included within the framework of quasi-degenerate perturbation theory. To accomplish this, the MRCI module of the ORCA program has been used, together with an effective one-electron mean field spin-orbit coupling operator \cite{SOMF}.
Further details are given in the SI.

\section{Acknowledgements} 
We thank J.\ I.\ Pascual, who was involved in the first experiments on Fe-OEP-Cl. We acknowledge discussions with C.\ Herrmann. We gratefully acknowledge funding by the Deutsche Forschungsgemeinschaft through Sfb 658 and HE7368/2, as well as the European Research Council through the consolidator grant "NanoSpin".

\clearpage

\setcounter{figure}{0}
\setcounter{section}{0}
\setcounter{equation}{0}
\renewcommand{\theequation}{S\arabic{equation}}
\renewcommand{\thefigure}{S\arabic{figure}}


\renewcommand{\Fig}[1]{\mbox{Fig.\unitspace\ref{Sfig:#1}}}
\renewcommand{\Figure}[1]{\mbox{Figure\unitspace\ref{Sfig:#1}}}

\section*{\Large{Supplementary Information}}

\section{Additional experimental data}

\subsection{Tip approach on different Fe-OEP-Cl molecules with different Pb tips}
In the main manuscript, we investigate the influence of the tip of the STM on the spin and oxidation state of Fe-OEP-Cl molecules at different tip-sample distances. In order to check the reproducibility of the experiments, different microscopic and macroscopic Pb tips as well as different molecules on different preparations were tested. We observe a reversible transition from regime I to regime II with all Pb tips and on all Fe-OEP-Cl molecules tested, albeit at  slightly different junction conductances. The stability in regime II and thus the extent of this regime differs for different tips. 
Only those experimental series, for which we do not observe any difference in the topography and $dI/dV$ spectra under standard tunneling conditions ($I=200\,$pA, $U=50\,$mV) before and after the tip approach are used for evaluation and presented in the manuscript and Supplementary Information.

\begin{figure*}[h]
	\includegraphics [width=0.95\textwidth,clip=]{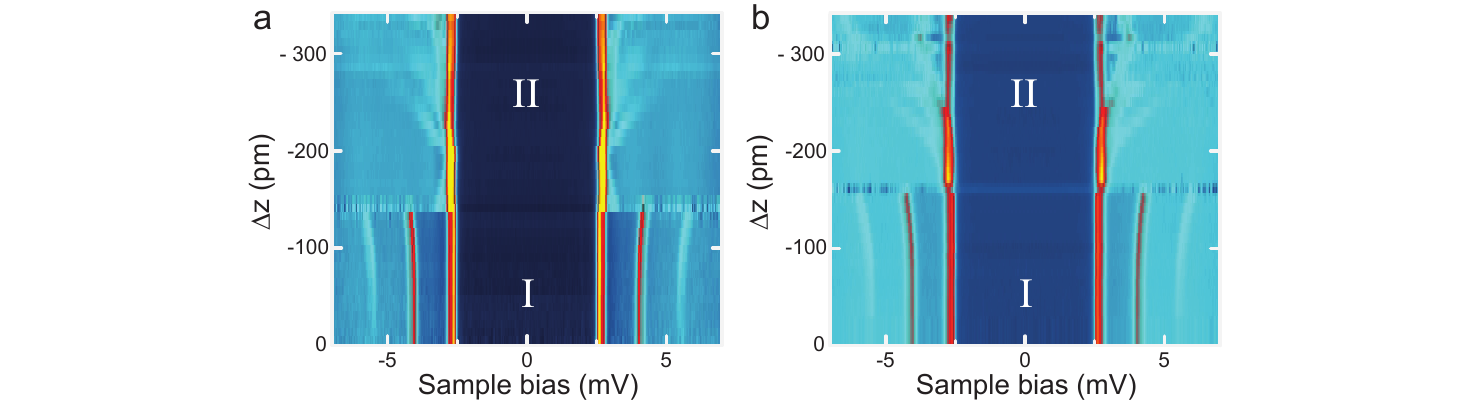}
	\caption{Pseudo-2D color plot distance-dependent $dI/dV$ spectra for two different Fe-OEP-Cl molecules on Pb(111). Both junctions are stable to at least $\Delta z=-360$\,pm. Set point: $V=50\,$mV; $I=200$\,pA.}
	\label{approachExamples}
\end{figure*}

Figure~\ref{approachExamples} presents two additional experimental series acquired with different tips and on different molecules. 
While the different experiments show qualitative agreement, there are differences with respect to the relative tip-sample distance $\Delta z$, for which the transition between regime I and regime II occurs. In Fig.~\ref{approachExamples}a, the transition occurs at $\Delta z\approx -150\,$pm, while it happens at $\Delta z\approx -170\,$pm in Fig.~\ref{approachExamples}b (starting from the same tunneling set point).
Furthermore, the evolution of the spin excitation patterns within regime II differs. For all molecules, we first observe a low-energy excitation, which shifts towards the superconducting gap edge. After this minimum of excitation energy, two excitations shift away from the superconducting gap edge. The shift of these excitations differs between the molecules. In some cases, a faint third excitation is observed. We ascribe the different excitation energies to small differences in the adsorption sites which are present within the island~\cite{Heinrich2015Ani_SI} and modify the crystal field of the Fe ion. Furthermore, different tip geometries alter the potential acting on the molecule during the tip approach. However, all observations agree with the $S=2$ state and the composition of the mixed spin projection states as described in the main text and in Fig.~4b of the main text. 

\subsection{Comparison of Fe-OEP with Fe-OEP-Cl in regime II}

\begin{figure}[h]
	\includegraphics [width=0.95\textwidth,clip=]{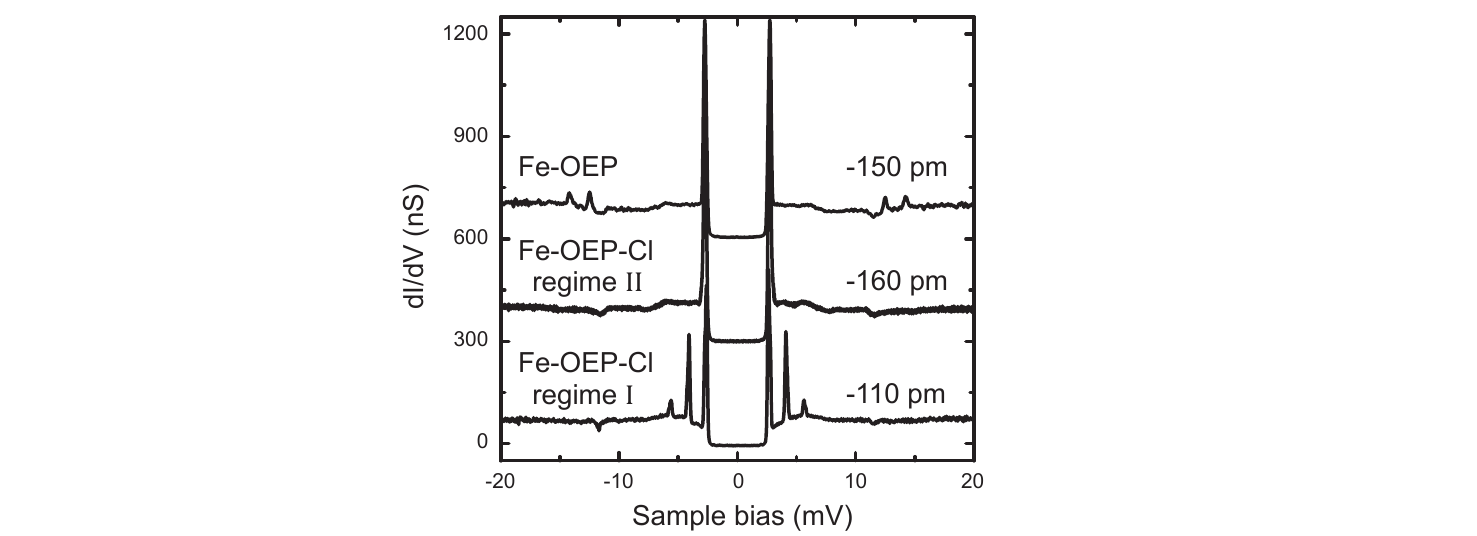}
	\caption{$dI/dV$ spectra  in a larger energy range of Fe-OEP, as well as of Fe-OEP-Cl in regime I and II, respectively. Set point: $V=50\,$mV; $I=200$\,pA. The spectrum of Fe-OEP shows two pairs of inelastic excitations at sample bias $V\approx \pm 12$mV and $V\approx \pm 14$\,mV, which are not observed for Fe-OEP-Cl, independent of the tip approach. }
	\label{long_range}
\end{figure}

Figure~\ref{long_range} compares $dI/dV$ spectra  of Fe-OEP with those of Fe-OEP-Cl in both regimes. There are two pairs of excitations at 
$\approx \pm 12$\,mV and $\approx \pm 14$\,mV observed in the case of Fe-OEP. These show small variations depending on the adsorption site of the molecule. Furthermore, they vary in energy with the tip-sample distance, but variations are restricted to less then $10\%$~\cite{Heinrich2015Ani_SI}.  For Fe-OEP-Cl, we do not observe  excitations with energies $>7\,$meV, which holds true for all junction configurations and reversible tip approaches. This is a further proof that
regime II of the Fe-OEP-Cl junction is not equivalent with a Fe-OEP junction. The Cl ligand is still present in the junction and actively influences the potential landscape.

\subsection{Tip approach in the normal state}

\begin{figure*}[h]
	\includegraphics [width=0.95\textwidth,clip=]{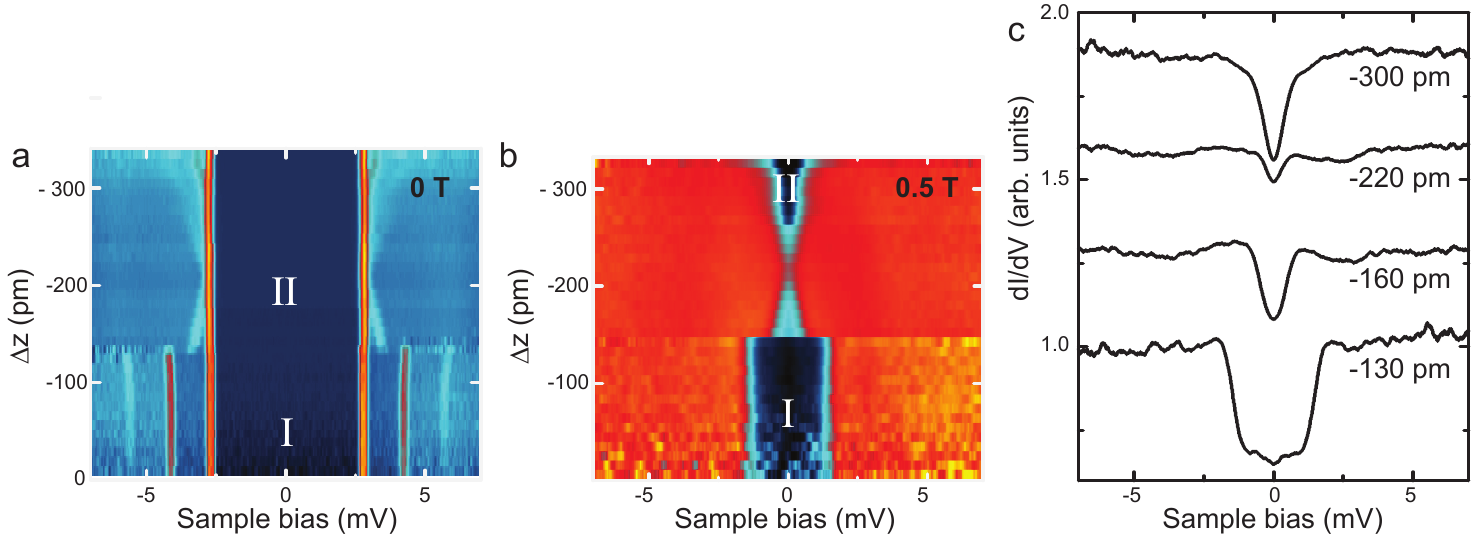}
	\caption{Pseudo-2D color plot of distance-dependent $dI/dV$ spectra on Fe-OEP-Cl on Pb(111) at 0\,T (a) and 0.5\,T (b), respectively. Same molecule as in Fig.~2a of the main manuscript. In (b), tip and substrate are in the normal, non-superconducting state. (c) Examples of $dI/dV$ spectra  as presented in (b) at various distances.  Set point: $V=50\,$mV; $I=200$\,pA.}
	\label{approach-B-field}
\end{figure*}

Figure~\ref{approach-B-field} presents a comparison of the approach experiment for tip and sample being in the superconducting (Fig.~\ref{approach-B-field}a) and in the normal state (Fig.~\ref{approach-B-field}b and c).  
Similar to the observations in the superconducting state, the inelastic excitations also shift in energy when approaching with the STM tip in the normal state. Because of the reduced energy resolution in the normal metal state, which is caused by the temperature-induced Fermi-Dirac broadening, these shifts are less well resolved compared to the experiments in the superconducting state. Yet, there is a qualitative agreement between both experiments. After the transition into regime II at $\Delta z\approx -150\,$pm, the excitation gap in the normal-state case closes with decreasing tip-sample distance, which corresponds to the excitation peak moving closer to the BCS coherence peaks in the superconducting state. Then, for $\Delta z < -220\,$pm, the gap widens again in the normal state, in agreement with the superconducting case, where excitations move away from the coherence peaks to higher energies. 

Note that in regime I the second pair of excitations is absent in the normal state, but present in the superconducting state. This is due to the increased lifetime in the latter~\cite{Heinrich2013_SI}, which allows for spin pumping.

\subsection{Zeeman shift of inelastic excitations}

\begin{figure*} []
	\includegraphics [width=0.95\textwidth,clip=]{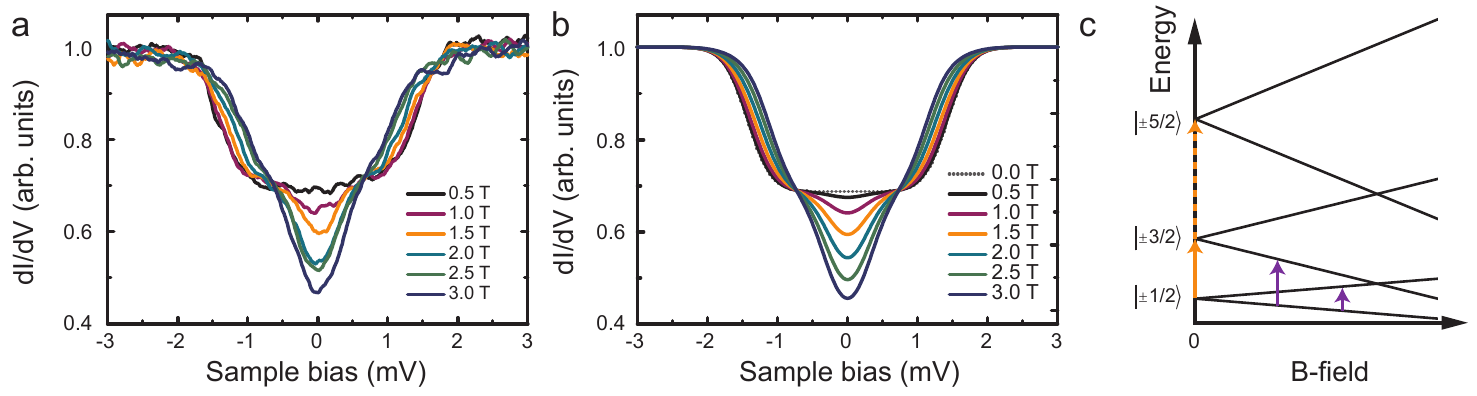}
	\caption{(a) B-field dependent $dI/dV$ spectra on Fe-OEP-Cl in regime I ($S=5/2$) at a constant tip-sample distance. Set point: $V=50\,$mV; $I=200$\,pA; $\Delta z = -100 $\,pm. (b) Simulated excitation spectra using $D=0.72$\,meV, $E=0$\,meV, $g=1.8$, and an effective temperature $T_\mathrm{eff}=1.3$\,K. (c) B-field dependent excitation diagram for a $S=5/2$ spin system with in-plane anisotropy and the B field oriented parallel to the anisotropy axis.  Adapted from Supporting Information of Ref.~\cite{Heinrich2015Ani_SI}.}.
	\label{Zeeman-S-2.5}
\end{figure*}

In the following, we discuss the Zeeman shift of the spin eigenstates in Fe-OEP-Cl. Figure~\ref{Zeeman-S-2.5}a shows a B-field-dependent measurement of the excitation spectrum in the normal state in regime I at constant distance. The B-field is oriented in $z$ direction. With increasing field strength, the excitation steps at approximately $\pm1.5\,$mV at $0.5$\,T move to lower energies with increasing field. 
Additionally, a dip appears at the Fermi energy $E_\mathrm{F}$ and gains weight with field strength. The excitation spectra can be simulated using an effective spin Hamiltonian model~\cite{Atanasov2015_SI}:
 \begin{equation}
	\hat{H_S} = g\mu_B \hat{S} \underline{B}+ D\hat{S_z^2}+ E(\hat{S_x^2}-\hat{S_y^2})
	 \quad . \label{spinHam}
\end{equation}
The first term accounts for the Zeeman splitting (with $g$ being the Land\'e g-factor, $\mu_B$ the Bohr magneton, \underline{B} the magnetic field, and $\hat{S}=(\hat{S_x}, \hat{S_y}, \hat{S_z})$ the spin operator). The second term describes the zero-field splitting in the presence of axial anisotropy with parameter $D$, while the last term also takes transverse anisotropy into account with the rhombic anisotropy parameter $E$. Using a temperature broadened Fermi-Dirac distribution, we can simulate the inelastic excitation spectra of the $S=5/2$ spin state with in-plane anisotropy ($D=0.72$\,meV) in B-fields of various strength parallel to the anisotropy axis. The simulation is done using the code by M. Ternes~\cite{Ternes2015_SI}. The result shown in Fig.~\ref{Zeeman-S-2.5}b  is in excellent agreement with the measurements presented in Fig.~\ref{Zeeman-S-2.5}a.
Figure~\ref{Zeeman-S-2.5}c shows a scheme of the corresponding energy levels of the spin eigenstates of the $S=5/2$ system with in-plane anisotropy (B parallel to the anisotropy axis). Because of the selection rules for inelastic spin excitations ($\Delta M_s=\pm 1; 0$), the only ground state excitation possible  at zero field is $M_s=\pm 1/2 \rightarrow \pm 3/2$.
At $B\neq 0$, a second excitation becomes possible ($M_s=-1/2 \rightarrow +1/2$), which increases in energy with $B$. It gives rise to the dip at $E_\mathrm{F}$ in the excitation spectrum because the limited energy resolution at 1.1\,K does not allow to resolve the step-like features (see also Theory section).

\begin{figure}[ht]
	\includegraphics [width=0.3\textwidth,clip=]{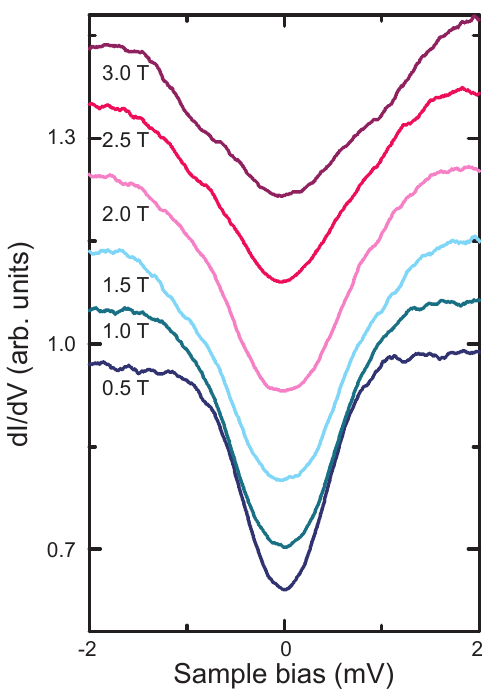}
	\caption{B-field dependent spectra at $\Delta z=-340$\,pm (same molecule as in Fig.~2a and Fig.~3 in the main text). Set point: $U=50\,$mV; $I=200$\,pA.}
	\label{Zeeman-S-2}
\end{figure}

In regime II, we also test for $B$-field-dependent excitations (Fig.~3 of the main manuscript). Here, we present a series of excitation spectra at different field strength at a smaller tip-sample separation ($\Delta z=-340$\,pm, Fig.~\ref{Zeeman-S-2}). Again, a $B$-field dependence is observed, the excitation step becoming wider with increasing $B$-field. This is most likely due to the shifting of two or more super-positioned steps, which cannot be resolved due to the thermal limitation in the energy resolution. Hence, we prove the magnetic origin of the excitation. 
\\

\section{Theory}
\subsection{Models}
We considered free molecules as model compounds for our quantum chemical investigations. Fe-OEP-Cl was used in experiments. As shown below, the ethyl groups do not have a significant effect on the crystal field of the Fe center, and, hence, on the magnetic properties. Therefore, we used Fe-P-Cl, where ``P'' stands for porphin, the simplest porphyrine 
 with eight peripheral H atoms instead of the ethyl groups, as a model system.

 For the Fe-P-Cl molecules, a neutral variant (with Fe in oxidation state +3) and an 
 anionic variant (with Fe in oxidation state +2) were considered.
 The chosen coordinate system is such 
 the Fe-Cl bond is aligned along the $z$-axis.
 
\subsection{Computational methodology}
As mentioned also in the Methods section of the main manuscript, the ORCA 
 (version 3.0.3) program package \cite{Neese2011_SI} was 
 used to optimize geometries and to calculate potential energy scans along 
 the Fe-Cl distance for several spin multiplicities. 
 These are the $S=0$, $S=1$, and $S=2$ multiplicities 
 for the anionic system, and $S=1/2$, $S=3/2$ and $S=5/2$ for the 
 neutral molecular model.
 For the scans, Fe-Cl distances in  a range between 
 2.1 and 3.2 \ \AA \ (in steps of 0.05  \AA) 
 were considered.

 First of all, the scans were done on the 
 DFT level of theory, utilizing 
 the B3LYP density functional \cite{Becke1993_SI} in combination with the def2-TZVP basis set \cite{Weigend2005_SI}.
To speed up the calculations, the RIJCOSX approximation \cite{RIJCOSX_SI} (in combination with the  def2-TZVP/JK auxiliary basis set) was used. 

For the calculation of spin-orbit coupling, zero-field splitted (excited) states, 
 but also for corrected potential energy curves of a given spin multiplicity, single-point 
 multi-reference WFT methods were used, following 
 established protocols \cite{Stavretis2015_SI,Atanasov2015_SI,Ganyushin2006_SI}. 
 For this purpose,  
 Complete Active Space Self Consistent Field (CASSCF) calculations 
  \cite{Malmqvist1989_SI} were performed.
 In all calculations, the orbitals were optimized in a state-averaged fashion and the quasi-restricted orbitals of the DFT calculations were used as guess orbitals.
In order to take into account effects of dynamical correlation, the NEVPT2 \cite{Angeli2001_SI,Angeli2001a_SI,Angeli2002_SI,Angeli2006_SI} method has been used on top of the CASSCF results. 
 The number of calculated states with respect to the given spin quantum number, $N(S)$,
  is, for the most important spin multiplicities as follows: $N(5/2)=1, N(3/2)=24, N(2)=5, N(1)=45$.

The effects of 
 spin-spin and spin-orbit interactions 
 included in the 
 form of quasi-degenerated perturbation theory using the following Hamiltonian:
\begin{equation}
	\hat{H} = \hat{H}_{\mathrm{0}} + \hat{H}_{\mathrm{soc}} + \hat{H}_{\mathrm{sp-sp}} + \hat{H}_{\mathrm{Ze}}
 \quad .
\label{exact}
\end{equation}
 Here, 
 $\hat{H}_{\mathrm{0}}$ is the usual, Born-Oppenheimer electronic 
 Hamiltonian, $\hat{H}_{\mathrm{soc}}$ is the spin-orbit coupling operator, approximated by an effective one electron spin-orbit mean field operator (SOMF) \cite{SOMF_SI}, $\hat{H}_{\mathrm{sp-sp}}$ accounts for spin-spin interaction and $\hat{H}_{\mathrm{Ze}} = \frac{1}{2}\mu_\mathrm{B} g_\mathrm{e}\left(\hat{L} + 2\hat{S} \right) \underline{B}$  is the Zeeman term which describes magnetic field interaction  ($\mu_\mathrm{B}$ is the Bohr magneton, $g_\mathrm{e}$ is the g-factor of the free electron, $\hat{L}$ and $\hat{S}$ the total angular and spin moment operators and $\underline{B}$ represents the magnetic field)\cite{Atanasov2012_SI}.
 To accomplish this, the MRCI module of the ORCA program has been used.  
 Within the CASSCF/NEVPT2/MRCI calculations the RI approximation (with the def2-TZVP/C auxiliary basis set) has been used \cite{RI-SSC_SI,RI-AUX_SI}.

For systems with spin $S\ge1$, 
  zero-field splitting 
 leads to splitting of spin states that are otherwise 
 degenerate.
 For analyzing this 
 effect in a simplified manner, effective spin Hamiltonians 
 are often used.  
 The Effective Hamiltonian method \cite{Ganyushin2006_SI,Stavretis2015_SI} is used to construct the
 spin Hamiltonian for a proper orientation (see discussion above and Eq.~\ref{spinHam}).  The eigenvalues of this Hamiltonian reproduce the splitting 
 of a spin ground state, into a $(2S+1)$-fold multiplet.
For example, neglecting the $E$-term, 
 the $S=2$ state (of, e.g., a Fe$^{2+}$ ion), splits 
 into states with $M_s=0$, $M_s=\pm 1$, and 
 $M_s=\pm 2$, separated by $D$ and $3D$, respectively. Likewise, 
 the $S=5/2$ state (of, e.g., a Fe$^{3+}$ ion), splits
 into states with $M_s=\pm 1/2$, $M_s=\pm 3/2$,
 and $M_s=\pm 5/2$, separated by $2D$ and $4D$, respectively.
 While this splitting behavior can be that simple (see below, for $S=5/2$), 
 more complicated cases arise for E$>$0 and involve a mixing of
 $M_s$ states (see main text for $S=2$, Fig.~4b).

We analyzed the character of 2S+1 states in Fig.~4b of the main text by calculating an averaged spin-projection eigenvalue along the $z$-axis:
\begin{equation}
	\bar{M}_s = \frac{1}{2S+1} \sum^{2S+1}_i |w_i\cdot M_{s,i}| 
\end{equation}
where, $w_i$ is the weight of the eigenstate of the spin projection operator along the $z$-axis with the eigenvalue $M_{s,i}$. $\bar{M}_s$ is calculated for all 2S+1 states. 

\subsection{Effect of neglecting the ethyl groups}
\label{without}
\begin{figure}[h]
	\begin{minipage}[t]{0.2\linewidth}
			\hspace{-3cm}
		\includegraphics[width=\linewidth]{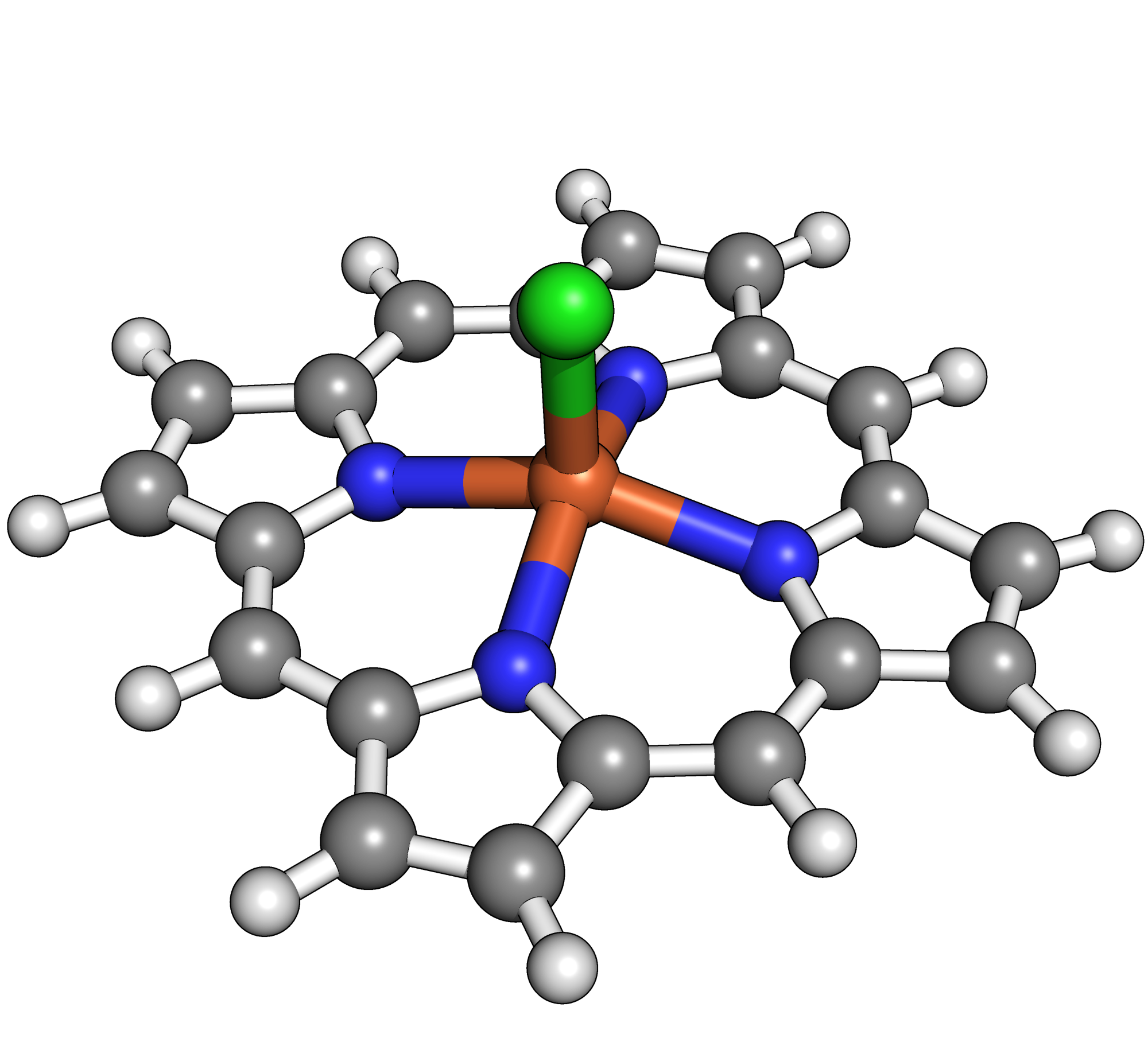}
	\end{minipage}	
	\begin{minipage}[t]{0.4\linewidth}
		\vspace{-3cm}
		\hspace{-3cm}
		\begin{tabular}{rl|ccc}
			                          &                      & Fe-OEP-Cl & Fe-P-Cl  & $\Delta$ \\ \hline
			                 r(Fe-Cl) & [\AA]                &  2.2335   &  2.2250  &  0.0085  \\
			                        D & $\mathrm{[meV]}$ &  0.151 & 0.153 & 0.002\\
			                      E/D &                      & 0.001505  & 0.001739 & 0.000234 \\
			$\mathrm{E(M_s=\pm 3/2)}$ & $\mathrm{[meV]}$ &    0.303    &  0.307 &   0.004  \\
			$\mathrm{E(M_s=\pm 5/2)}$ & $\mathrm{[meV]}$ &    0.908    &  0.920 &   0.012
		\end{tabular}
	\end{minipage}
	\caption{Left: Optimized structure of the Fe-P-Cl model. Right: Comparison between geometrical and anisotropy parameters. The deviations, $\Delta$, 
  between both models are rather small. As a consequence, the computational less demanding model (Fe-P-Cl) has been used in our study.
}
	\label{fig:figs6fepcl}
\end{figure}

To make sure that our simplified Fe-P-Cl model system
is able to represent the characteristics of the Fe-OEP-Cl system, 
 a test calculation for the 
 neutral compound 
 at its equilibrium geometry has been performed. 
 We compare the resulting Fe-Cl bond length as well as the energies of the magnetic states 
 (and the anisotropy parameters). The results are shown together with the geometry of the Fe-P-Cl complex in Fig.~\ref{fig:figs6fepcl}. We conclude the ethyl side groups have a negligible effect on the electronic and magnetic properties of the system. This is the reason why in all other calculations 
 presented in this work, the simplified and computationally less demanding, 
 Fe-P-Cl model system has been used.

\subsection{Potential energy curves at the B3LYP level of theory}
\begin{figure}[hbt]
  \includegraphics [width=0.3\textwidth,clip=]{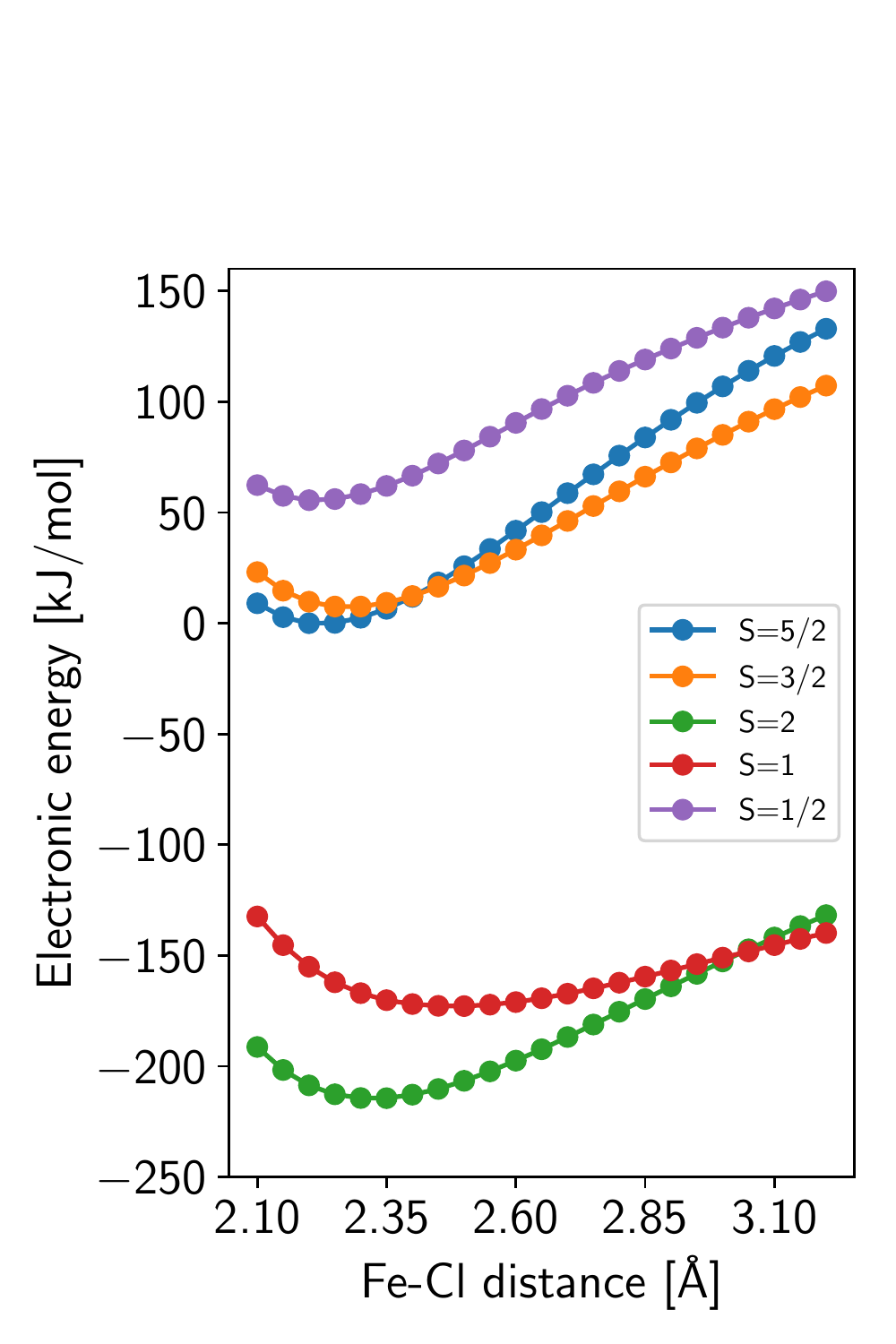}
  \caption{Potential energy curves for various spin states 
 of neutral Fe-P-Cl (Fe$^{3+}$, $S=1/2, 3/2, 5/2$) 
 and anionic  Fe-P-Cl  (Fe$^{2+}$, $S=1, 2$; $S=0$ is at  high energy and not shown), 
 calculated with B3LYP/def2-TZVP.
}
\label{fig:dft_surface}
\end{figure}

As already stated in the main text, geometries were 
 optimized at selected, 
  fixed Fe-Cl distances with the B3LYP method. The electronic energies of the DFT calculations as a function of the Fe-Cl distance are 
 plotted in Fig.~\ref{fig:dft_surface}. Since the relative 
 stability of low- {\em vs.} high-spin 
 states are highly dependent on the amount of added ``exact'' (Hartree-Fock-like) exchange in the DFT functional \cite{DFT-EXC_SI}, 
 we prefer to show the NEVPT2  energies (at B3LYP geometries) in the main text instead (Fig.~4a there).
 The main conclusions of this work are unaffected by this choice.

\subsection{Zero-field splittings and anisotropy parameters for $S=5/2$}
\begin{figure}[h]
  \includegraphics [width=0.6\textwidth,clip=]{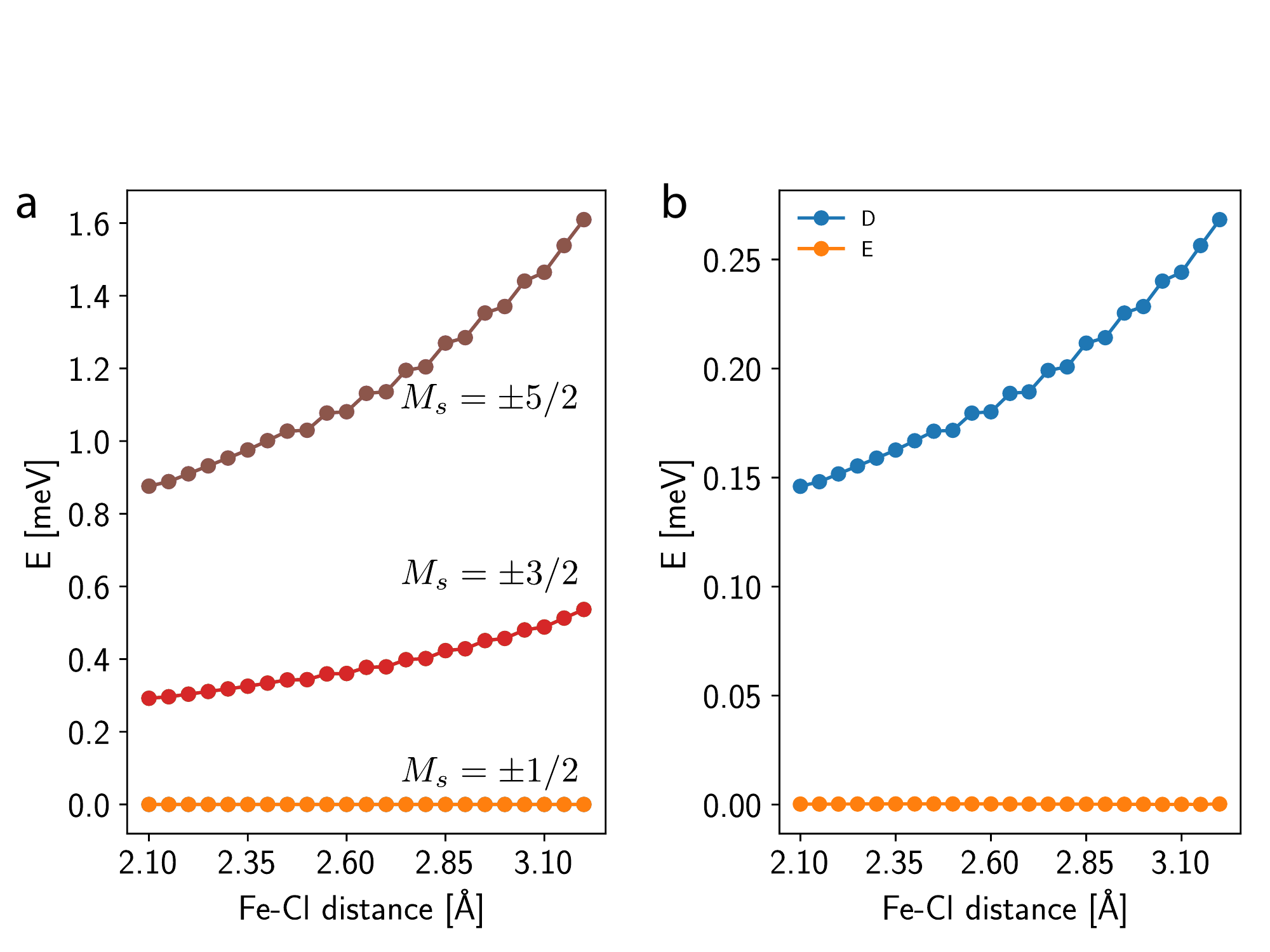}
  \caption{(a) Zero-field splitted 
 states as a function of Fe-Cl distance for the neutral, 
 $S=5/2$ system calculated using the CASSCF/NEVPT2/MRCI methodology. (b) Anisotropic ZFS 
 $D$ and $E$ parameters extracted with the effective Hamiltonian method.}
\label{fig:mrci_soc_6}
\end{figure}

At all points along the potential energy surface scan, the zero-field splitting 
 was calculated. The results (state energies and anisotropy parameters) for the 
 $S=5/2$ state of the neutral (Fe$^{3+}$) complex are shown 
 in Fig.~\ref{fig:mrci_soc_6}. 

For the state energies in Fig.~\ref{fig:mrci_soc_6}a, 
 we observe a splitting of the six-fold degenerate ground state $S=5/2$ into three Kramer doublets 
 with $M_s=\pm 1/2, \pm 3/2$ and $\pm 5/2$, respectively,  
 as expected, when spin-orbit coupling and the spin-spin interaction are considered. At the equilibrium Fe-Cl distance of 
 2.20 \AA,  the 
 computed state energies, relative 
 to the ground state, are 
  $\mathrm{E}_1 = 0.3~\mathrm{meV}$ and $\mathrm{E}_2 = 0.91~\mathrm{meV}$. 
 Consequently, the excitation energies are $\epsilon_1 = 0.3~\mathrm{meV}$ ($\mathrm{E}_1$-$\mathrm{E}_0$) 
 and $\epsilon_2 = 0.61~\mathrm{meV}$ ($\mathrm{E}_2$-$\mathrm{E}_1$), respectively.
 These excitation energies are  lower than in the experiment (approx. 1/5 of the experimental value). 
Since lower calculated excitation energies have been observed also for other 
 iron(III) porphyrin halides by Stavretis \textit{et al.} \cite{Stavretis2015_SI} who used a 
 very similar method as we do, this seems to be a systematic effect. 

Further, we see that 
 with increasing Fe-Cl bond length, the 
  excitation energies become larger. This is in agreement with experiment, see Fig.~2a of the main text, for 
 the region $ -120~\mathrm{pm} \leq \Delta z \leq 0~\mathrm{pm} $. 

In Fig.~\ref{fig:mrci_soc_6}b, we show the ZFS anisotropy parameters $D$ and $E$ derived 
 from the model Hamiltonian (Eq.~\ref{spinHam}) by comparing to the state energies calculated from Eq.~\ref{exact}, as a function of Fe-Cl distance. 
(For instance, 
 according to the model Hamiltonian we have 
 $\epsilon_1 = \mathrm{E}_1$-$\mathrm{E}_0 \sim 2 D$ and 
 $\epsilon_2 = \mathrm{E}_2$-$\mathrm{E}_1 \sim 4 D$, 
 see above.)
 At around the 
 equilibrium bond length, we find $D \sim +0.15$ meV and 
 $ E \sim 0$. $D$ increases with bond length.

\subsection{Anisotropy parameters for $S=2$}
\begin{figure}[h]
	\includegraphics [width=0.3\textwidth,clip=]{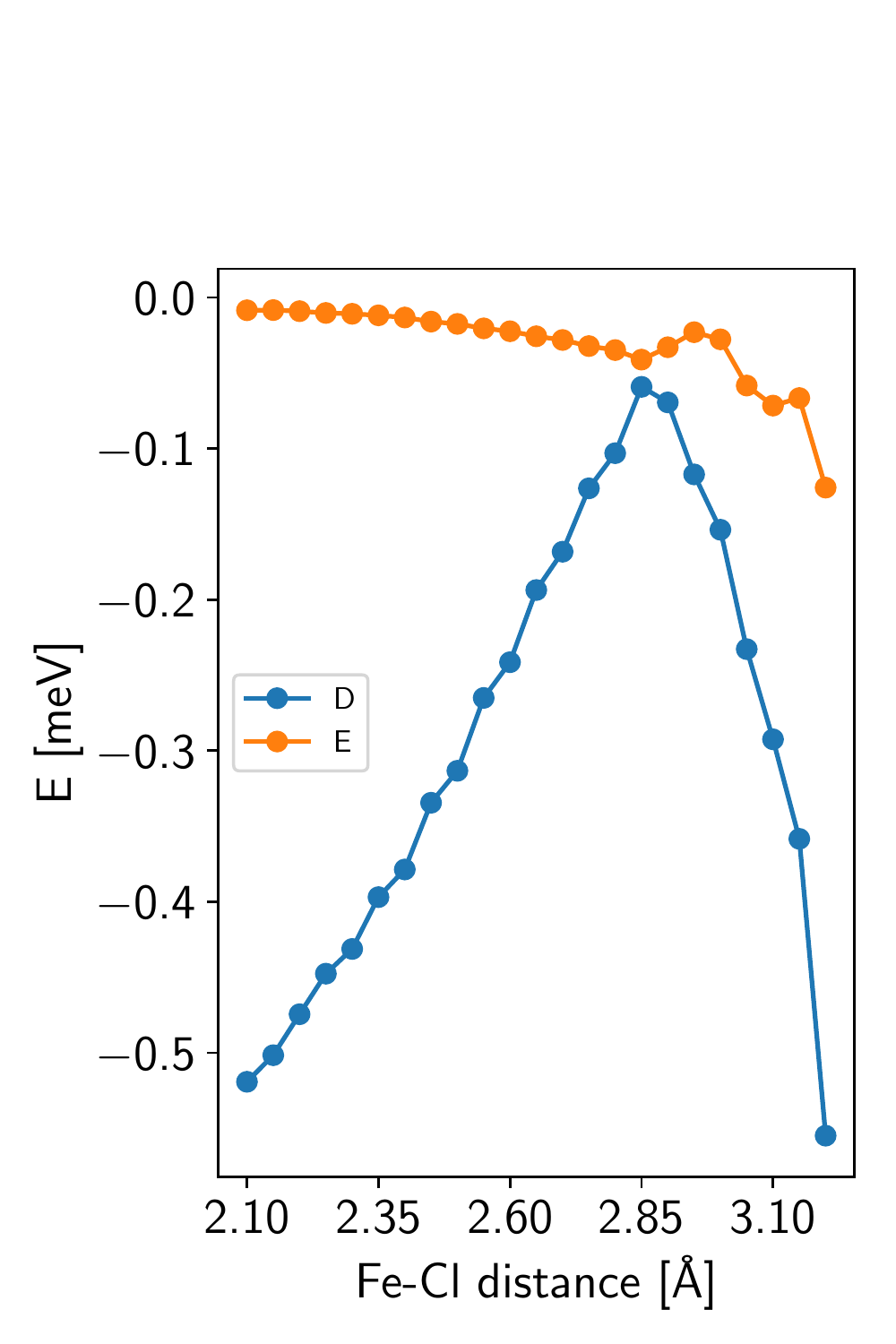}
	\caption{Anisotropic ZFS parameters $D$ and $E$ as a function of Fe-Cl distance, 
 calculated for the anionic $S=2$ manifold using the CASSCF/NEVPT2/MRCI methodology. }
	\label{fig:de_5}
\end{figure}

We also extracted the anisotropy parameters $D$ and $E$ for the anionic ground state, 
 $S=2$. This is shown in Fig.~\ref{fig:de_5}. In this case 
 $D$ and $E$ values show 
 a more complex behavior, and both being negative. 
 However, we should also note that 
 in this case there is some ambiguity when determining 
 parameters, since different data points belong to different orientations of the anisotropy axes. 
 
\subsection{Interaction with a magnetic field}
\begin{figure}[h]
  \includegraphics [width=0.3\textwidth,clip=]{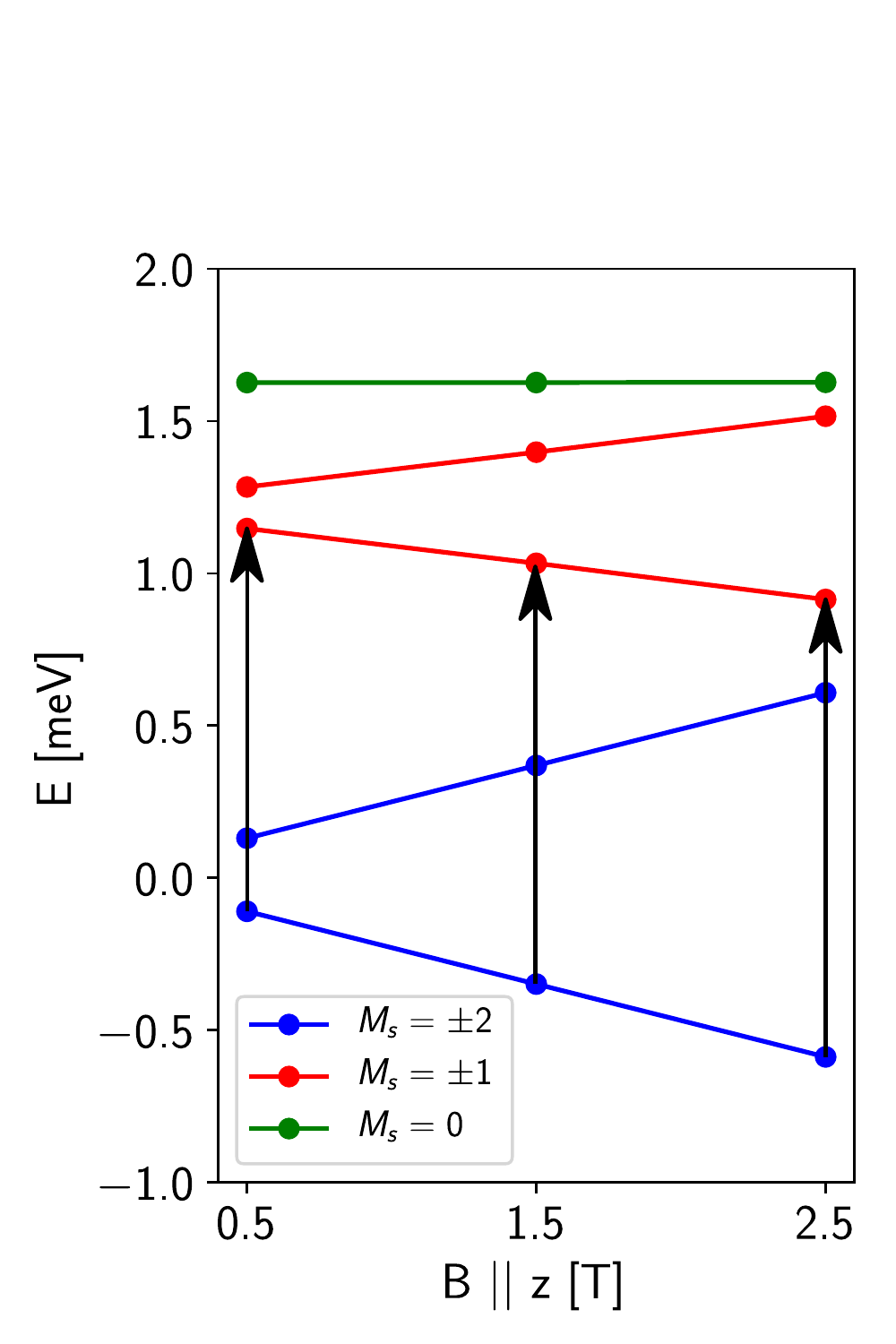}
  \caption{Ground and excited states of the anionic (Fe$^{2+}$) Fe-P-Cl model for 
 $S=2$, 
 at its equilibrium geometry. The state energies are shown as a function of strength of an applied magnetic field along the $z$-axis. Arrows indicate allowed excitations from the ground state.}
\label{fig:BFlield}
\end{figure}

In addition to the simulation of zero-field splittings, 
 the influence of an additional magnetic field $B$ has been investigated, oriented along 
 the $z$-axis (the Fe-Cl axis). In relation to the measurements presented in 
 Fig.~3a and b of the main text
  at a distance of $\Delta z$ = -150 pm, we performed simulations at the relaxed geometry of the negatively charged 
 Fe(II) complex with a spin quantum number of $S=2$. In  Fig.~\ref{fig:BFlield} we show 
 states arising from the ``exact'' Hamiltonian (\ref{exact}) including the Zeeman term.

 Since the degenerate ground states (no field) are composed of the eigenvectors of the spin projection operator ($z$-axis) with eigenvalues of $M_s=\pm 2$ (see Fig.~4a of the main text), the energies of these states are heavily affected by the external magnetic field. The third and fourth state ($M_s = \pm 1$) are less affected and the energy of the fifth state does not change at all ($M_s = 0$). Since only transitions between states with $\Delta M_s=0, \pm 1$ are allowed, an increasing excitation energy is the consequence (from the ground to the second excited state). This is in agreement with the experiment, where an increasing field-dependent excitation energy is observed. 

\end{document}